\documentclass[12pt,preprint]{aastex}

\usepackage{graphicx}

\shorttitle{EIS Moss Observations}
\shortauthors{Warren et al.}

\slugcomment{\texttt{Submitted to ApJ September 4, 2007}}

\begin{document}


\title{Observation and Modeling of Coronal ``Moss'' With the EUV
Imaging Spectrometer on \textit{Hinode}}

\author{Harry P. Warren\altaffilmark{1}, 
        Amy R. Winebarger\altaffilmark{2},
	John T. Mariska\altaffilmark{1}, 
	George A. Doschek\altaffilmark{1}, 
	Hirohisa Hara\altaffilmark{3}}

\altaffiltext{1}{Space Science Division, Naval Research Laboratory,
  Washington, DC 20375, \texttt{hwarren@nrl.navy.mil}}

\altaffiltext{2}{Department of Physics, Alabama A\&M, 4900 Meridian
Street, Normal, AL 35762}

\altaffiltext{3}{National Astronomical Observatory of Japan, National
 Institutes of Natural Sciences, Mitaka, Tokyo, 181-8588, Japan}


\begin{abstract}
 Observations of transition region emission in solar active regions
 represent a powerful tool for determining the properties of hot
 coronal loops. In this Letter we present the analysis of new
 observations of active region moss taken with the Extreme Ultraviolet
 Imaging Spectrometer (EIS) on the \textit{Hinode} mission. We find
 that the intensities predicted by steady, uniformly heated loop
 models are too intense relative to the observations, consistent with
 previous work. To bring the model into agreement with the
 observations a filling factor of about 16\% is required. Furthermore,
 our analysis indicates that the filling factor in the moss is
 nonuniform and varies inversely with the loop pressure.
\end{abstract}

\keywords{Sun: corona}


\section{Introduction}

 Understanding how the solar corona is heated to high temperatures is
 one of the most important problems in solar physics. In principle,
 observations of emission from the solar corona should reveal
 important characteristics of the coronal heating mechanism, such as
 the time scale and location of the energy release. In practice,
 however, relating solar observations to physical processes in the
 corona has proved to be very difficult. One of the more significant
 obstacles is the complexity of the solar atmosphere, which makes it
 difficult to isolate and study individual loops, particularly in the
 cores of solar active regions.

 One circumstance in which the line of sight confusion can be reduced
 is observations of coronal ``moss.'' The moss is the bright,
 reticulated pattern observed in many EUV images of solar active
 regions. These regions are the footpoints of high temperature active
 region loops, and are a potentially rich source of information on the
 conditions in high temperature coronal loops (see
 \citealt{peres1994,berger1999,fletcher1999,depontieu1999,martens2000}).

 A particularly useful property of the moss is that its intensity is
 proportional to the total pressure in the coronal loop
 \citep{martens2000,vourlidas2001}, which is an important constraint
 for coronal loop modeling. For steady heating models the pressure and
 the loop length completely determine the solution up to a filling
 factor (see, \citealt{rosner1978}). Recently \cite{winebarger2007}
 used this property of the moss in conjunction with a magnetic field
 extrapolation to infer the volumetric heating rate for each field
 line in the core of an active region.  This allowed them to simulate
 both the soft X-ray and EUV emission independent of any assumptions
 about the relationship between the volumetric heating rate and the
 magnetic field.

 One limitation of this approach is that the filling factor is left as
 a free parameter and is adjusted so that the soft X-ray and EUV
 emission best match the observations. Observations of both the
 intensity and the electron density in the moss would determine the
 filling factor and remove this degree of freedom. A more fundemental
 question is how relevant steady heating models are to coronal
 heating. Recent work has shown that high temperature coronal emission
 ($\sim 3$\,MK) can be modeled successfully with steady heating (e.g.,
 \citealt{schrijver2004,lundquist2004,warren2006b}). However,
 pervasive active region Doppler shifts (e.g.,
 \citealt{winebarger2002}) and the importance of non-equilibrium
 effects in lower temperature loops (e.g., \citealt{warren2003}),
 suggest an important role for dynamical heating in the solar corona.

 The purpose of this paper is to address these two issues with new
 observations from the Extreme Ultraviolet Imaging Spectrometer (EIS)
 on \textit{Hinode}. EIS has an unprecedented combination of high
 spatial, spectral, and temporal resolution and provides a unique view
 of the solar corona. Here we use EIS \ion{Fe}{12} density sensitive
 line ratios to infer the density and pressure of the moss and
 determine the filling factor of high temperature coronal loops. We
 also use EIS measurements of several emission lines formed near 1\,MK
 to test the consistency of steady heating models with moss observations.

\section{Modeling EIS Moss Observations}

 In this section we discuss the application of steady heating models
 to observations of the moss with EIS. To begin we calculate the
 plasma emissivites as a function of both temperature and density for
 ions relevant to EIS using the CHIANTI atomic physics database
 \citep[e.g.,][]{young2003}. We use the ionization fractions of
 \cite{mazzotta1998} and the coronal abundances of
 \cite{feldman1992}. The plasma emissivities are related to the
 observed intensity by the usual expression
 \begin{equation}
  I_\lambda = \frac{1}{4\pi}\int_s\epsilon_\lambda(n_e,T_e)n_e^2\,ds,
  \label{eq:ints}
 \end{equation}
 where $n_e$ and $T_e$ are the electron density and temperature and
 $s$ is a coordinate along the field line.  Note that what we refer to
 as the emissivity ($\epsilon_\lambda$) is actually the emissivity
 divided by $n_e^2$. For many strong emission lines this quantity is
 largely independent of the density. In this context, however,
 including the density dependence explicitly is important since many
 of the EIS emission lines have emissivities that are sensitive to the
 assumed density. Our calculations cover the range $\log\,n_e = 6$--13
 and $\log\,T_e=4$--8.
 
 To explore the variation of the observed emission with base pressure
 we have calculated a family of solutions to the hydrodynamic loop
 equations using a numerical code written by Aad van Ballegooijen
 (e.g., \citealt{schrijver2005}). We consider total loop lengths in
 the range $L = 10$--100\,Mm and maximum temperatures in the range
 $\log T_{max}=2.5$--7.5\,MK, which are typical of active region
 cores. The loops are assumed to be oriented perpendicular to the
 solar surface and to have constant cross sections. For consistency
 with our emissivity calculations we use the more recent radiative
 loss calculations given in \cite{brooks2006}.

\begin{figure*}[t!]
 \centerline{%
 \includegraphics[scale=0.565]{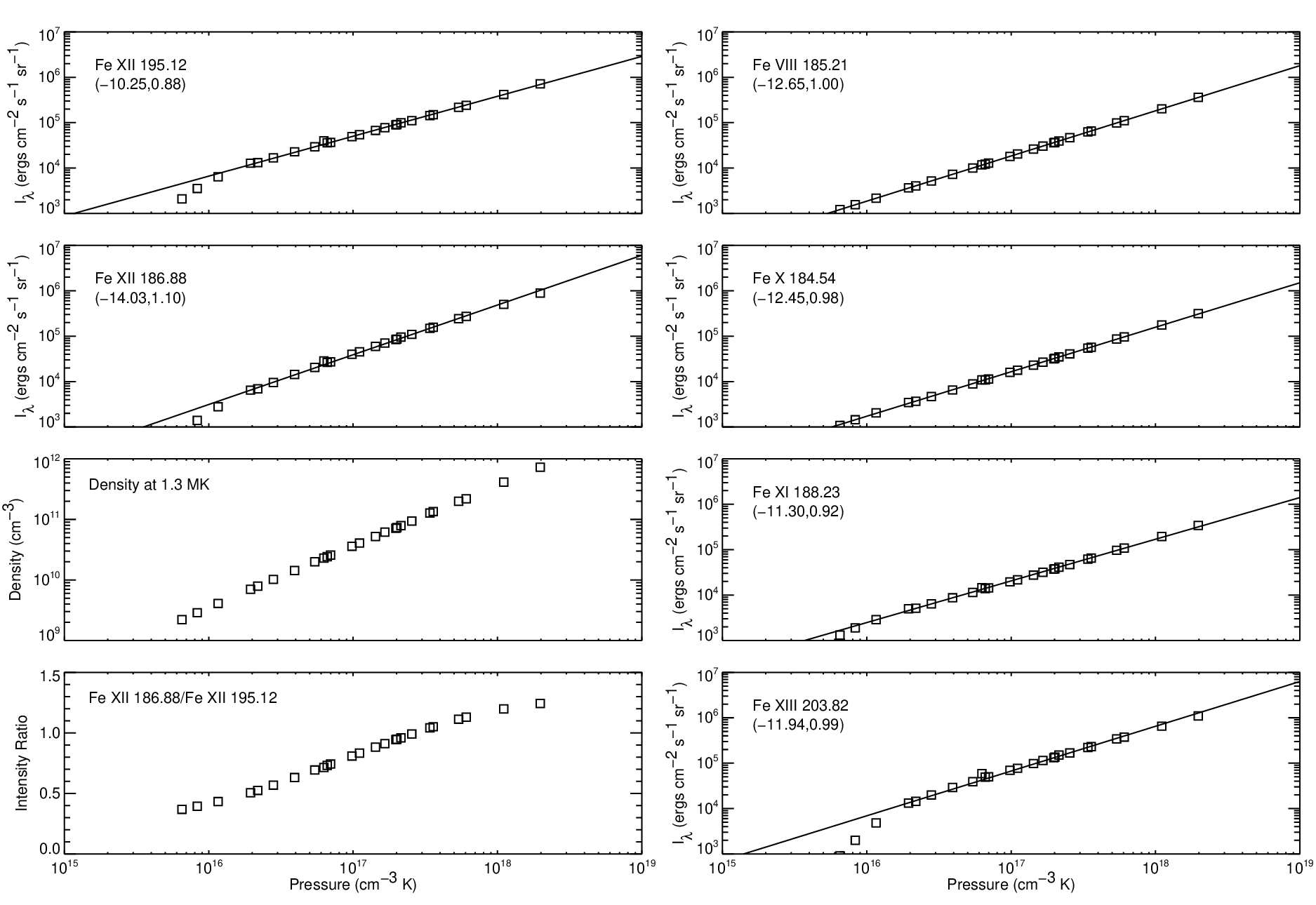}}
 \caption{Plots of line intensity as a function of base pressure for
 several emission lines that can be observed with EIS.  A filling
 factor of unity is assumed. The coefficients of the power-law fit
 ($\log_{10}a_ \lambda, b_\lambda$) are indicated. Also shown
 (\textit{bottom left panels}) are the density at 1.3\,MK, the peak
 temperature of formation for \ion{Fe}{12}, and the \ion{Fe}{12}
 186.88/195.12\,\AA\ ratio. The density and the density sensitive line ratio
 are independent of the filling factor and allow the pressure to be
 inferred from the observations.}
 \label{fig:calc}
 \end{figure*}

 Each solution to the loop equations gives the density and temperature
 as a function of position along the loop ($n_e(s),T_e(s)$). We
 interpolate to find the emissivity from the calculated density and
 temperature in each computational cell. We then integrate the
 emissivity at heights below 5\,Mm to determine the total footpoint
 intensity using Equation~\ref{eq:ints}. The resulting intensities are
 displayed in Figure~\ref{fig:calc} and are generally consistent with
 earlier work which showed that the moss intensities are proportional
 to the pressure and independent of the loop length
 \citep[e.g.,][]{martens2000,vourlidas2001}. The linear relationship
 between the intensity and the pressure breaks down for some of the
 hotter lines for the lowest pressures we've considered. For these
 solutions the peak temperature of formation for the line is closer to
 the apex temperature and there is some emission from along the loop
 leg.

 For each line we perform a fit of the form $I_\lambda = a_\lambda
 p_0^{b_\lambda}$ to the calculated intensities. Only pressures above
 $10^{16}$\,cm$^{-3}$ K are considered in the fits. The exponent, which
 is also shown in Figure~\ref{fig:calc}, is generally close to 1. One
 approximation that has been made in earlier theoretical work that is
 not strictly valid is that the emissivity is independent of the
 density. Our calculations account for this and are more accurate.

 These model calculations suggest the following recipe for deriving
 the loop pressure and filling factor from the observations. The
 \ion{Fe}{12} 186.88/195.12\,\AA\ ratio can be used to infer the
 pressure. The pressure then can be used to determine the expected
 intensities in each of the lines. The filling factor is derived from
 the ratio of the simulated to observed intensity.

\section{EIS Moss Observations}

 To test these model calculations we use observations from EIS on
 \textit{Hinode}, which was launched 23 September 2006. The EIS
 instrument produces high resolution solar spectra in the wavelength
 ranges of 170--210\,\AA\ and 250--290\,\AA. The instrument has
 1\arcsec\ spatial pixels and 0.0223\,\AA\ spectral pixels. Further
 details are given in \cite{culhane2007} and \cite{korendyke2006}.

 Despite the fact that \textit{Hinode} was launched close to the
 minimum in the solar activity cycle there have been several active
 regions available for observation.  For this work we analyzed
 \textit{Hinode} observations of NOAA active region 10940 from 2007
 February 2 at about 10:42 UT. These observations show areas of moss
 that are relatively free from contamination from other emission in
 the active region. The observing sequence for this period consisted
 of stepping the 1\arcsec\ slit over a $256\arcsec\times256\arcsec$
 region taking 15\,s exposures at each position. 

\begin{figure*}[t!]
 \centerline{%
 \includegraphics[scale=0.55]{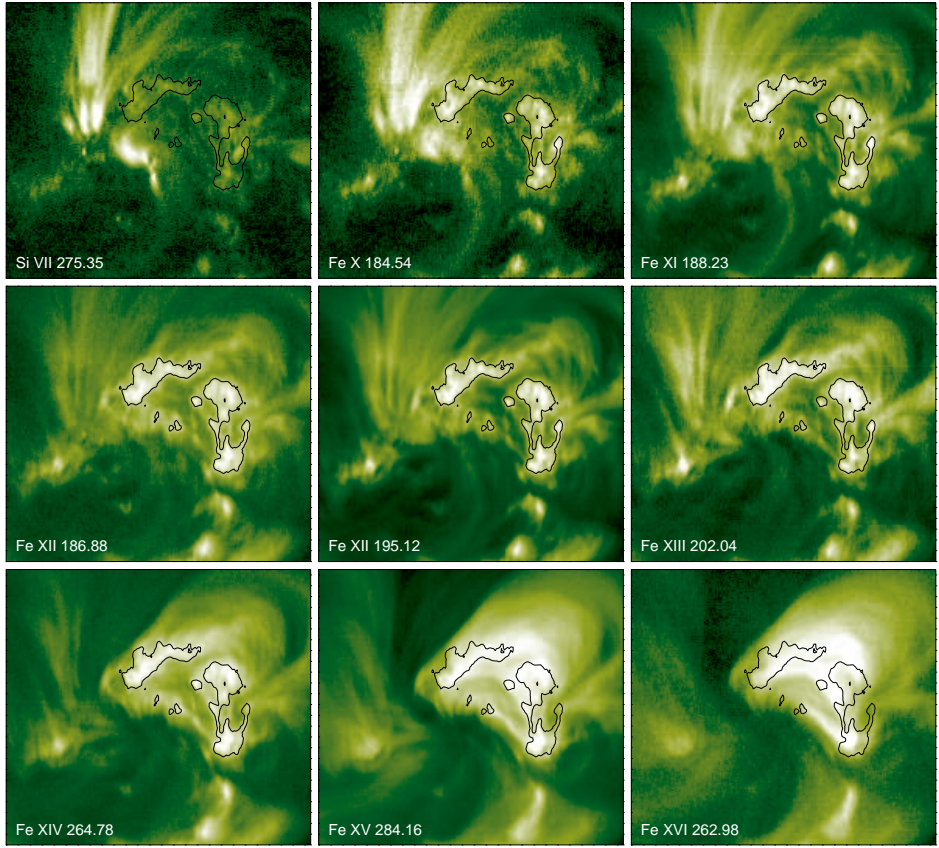}}
 \caption{EIS rasters of NOAA active region 10940 in 9 emission
 lines. The contour is derived from the \ion{Fe}{12} 186.88\,\AA\
 emission line and represents the approximate location of the moss. A
 subfield of the raster approximately $196\arcsec\times178\arcsec$ is
 shown. The emission from \ion{Fe}{14} and the lines formed at higher
 temperatures is more loop-like and not considered to be part of the
 active region moss.}
 \label{fig:rast}
 \end{figure*}

 These data were processed to remove the contribution of CCD
 background (pedestal and dark current), electron spikes, and hot
 pixels, and converted to physical units using standard software. We
 then fit the calibrated data assuming Gaussian profiles and a
 constant background. The resulting rasters for some of the strongest
 lines are shown in Figure~\ref{fig:rast}.

 To identify the moss in this region we adopt the very simple strategy
 of looking for the brightest \ion{Fe}{12} 186.88\,\AA\ emission. This
 line is very sensitive to density. The line intensity relative to
 \ion{Fe}{12} 195.12\,\AA\ rises by about a factor of 6 as the
 densities rises from $10^9$, which is typical of active region loops
 at this temperature, to $10^{11}$\,cm$^{-3}$, which is at the high
 end of what we expect to observe in the moss. The result of using a
 simple intensity threshold is shown in Figure~\ref{fig:rast}.

\begin{figure}[b!]
 \centerline{%
 \includegraphics[clip,scale=0.565]{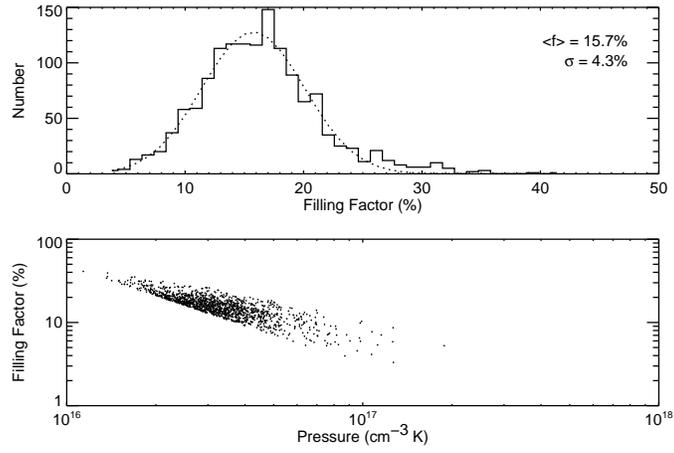}}
 \caption{Scatter plots of simulated versus observed intensity for
 moss observed with EIS. The simulated intensities include a filling
 factor derived from the \ion{Fe}{12} 195.12\,\AA\ line.}
 \label{fig:filling}
 \end{figure}

 For each of the 1416 spatial pixels identified as moss we use the
 \ion{Fe}{12} 186.88/195.12\,\AA\ ratio to determine the base pressure
 in the loop from the calculation shown in Figure~\ref{fig:calc}. The
 base pressure is then used to infer the intensities in all of the
 emission lines, again using the calculations of this type also shown
 in Figure~\ref{fig:calc}. As has been found in many previous studies,
 the simulated intensities are much higher than what is observed. To
 reconcile the observed and simulated intensities we introduce a
 filling factor for each point derived from the \ion{Fe}{12}
 195.12\,\AA\ intensities. As shown in Figure~\ref{fig:filling}, the
 distribution of filling factors is approximately Gaussian with a peak
 at about 16\% and standard deviation of about 4\%. The mean filling
 factor we derive here is generally consistent with previous results
 that have been about 10\% \citep[e.g.,][]{porter1995,martens2000}. We
 also find that the derived filling factor is nonuniform and inversely
 proportional to the base pressure.

\begin{figure*}[t!]
 \centerline{%
 \includegraphics[scale=0.565]{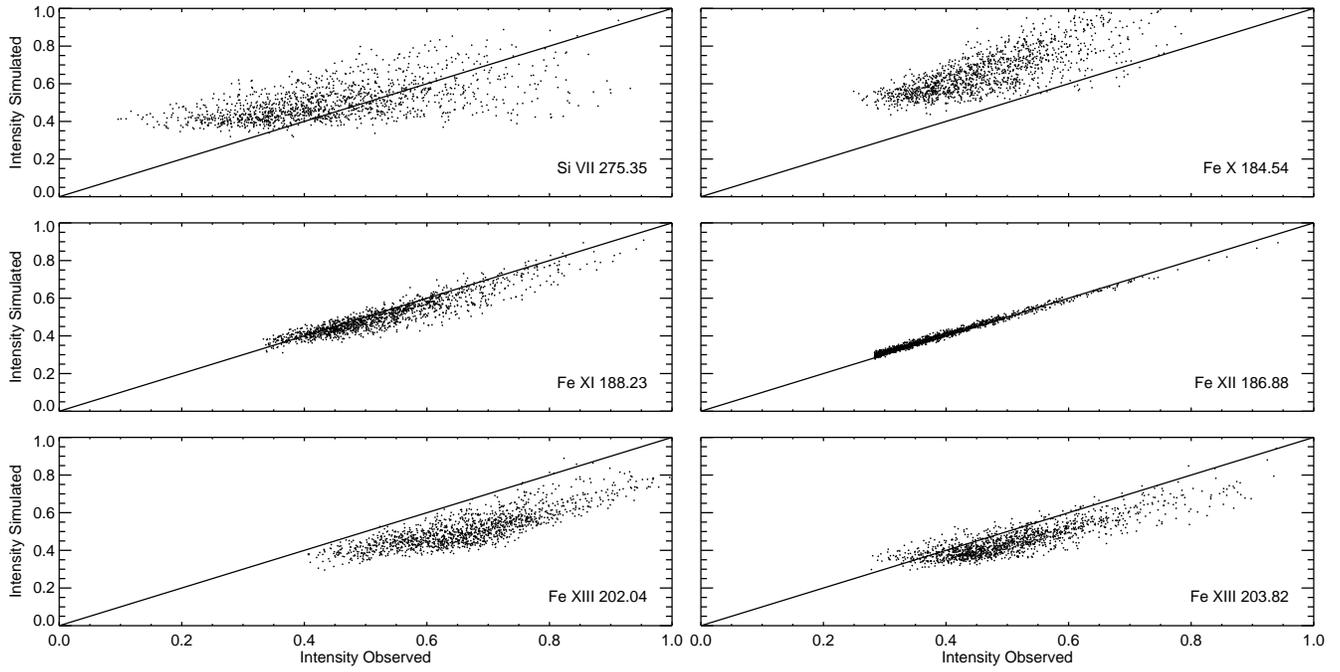}}
 \caption{Scatter plots of simulated versus observed intensity for
 moss observed with EIS. The simulated intensities include a filling
 factor derived from the \ion{Fe}{12} 195.12\,\AA\ line. The
 intensities in each emission line are normalized to the maximum
 observed intensity. An additional factor of 4 has been used to scale
 the \protect{\ion{Si}{7}} intensities.}
 \label{fig:compare}
 \end{figure*}

 To compare the model with the observation we have made scatter plots
 of observed intensity versus simulated intensity (including the
 filling factor) for a number of emission lines. These plots are shown
 in Figure~\ref{fig:compare} and indicate a reasonable agreement
 between the steady heating model and the observations. The agreement
 is particularly good for \ion{Fe}{11} 188.23 and \ion{Fe}{13}
 203.82. The simulated \ion{Fe}{10} 184.54\,\AA\ intensities are about
 20\% too high and the simulated \ion{Fe}{13} 202.04 are about 10\%
 too low. These discrepancies are not alarming considering the
 uncertainties in the atomic data.
 
 There are, however, rather signficant discrepancies in the
 \ion{Si}{7} 275.35\,\AA\ intensities. The simulated intensities are
 generally about a factor of 4 larger than what is predicted from the
 steady heating model. The correlation between the simulated and
 observed intensities is also poor. It is possible that considering
 loop expansion or nonuniform heating may provide a better match at
 these temperatures, but we have not yet performed these
 simulations. It is also possible that there are errors in the atomic
 data for this line or that the instrumental calibration is not
 correct at this wavelength. More extensive analysis, including the
 observation of other lines formed at similar temperatures, will be
 required to resolve this issue.

\section{Discussion}

 We have presented an initial analysis of active region moss observed
 with the EIS instrument on \textit{Hinode}. We find that the
 intensities predicted by steady, uniformly heated loop models are too
 high and a filling factor is required to bring the simulated
 intensities into agreement with the observations. The mean filling
 factor we derive here ($\sim16$\%) is similar to that determined from
 earlier work with steady heating models. Furthermore, we also find
 that the filling factor in the moss must be nonuniform and varies
 inversely with the loop pressure.

 The next step is to use the methodology outlined here to simulate all
 of the active region emission (including the high temperature loops)
 and compare with EIS observations. Finally, we emphasize that while
 we find reasonable agreement between the moss intensities and steady
 heating, we cannot rule out dynamical heating processes, such as
 nano-flares \citep[e.g.,][]{cargill1997,cargill2004}. The X-ray
 emission observed in this region clearly has a dynamic component
 \citep{warren2007b} and it would be surprising if non-equilibrium
 effects did not play some role in the heating of this active region.


\acknowledgments Hinode is a Japanese mission developed and launched
by ISAS/ JAXA, with NAOJ as domestic partner and NASA and STFC (UK) as
international partners. It is operated by these agencies in
co-operation with ESA and NSC (Norway).



\clearpage

\clearpage

\clearpage

\clearpage

\begin{thebibliography}{}

\bibitem[\protect\citeauthoryear{{Berger} et~al.}{{Berger}
  et~al.}{1999}]{berger1999}
{Berger}, T.~E., {de Pontieu}, B., {Schrijver}, C.~J.,  \& {Title}, A.~M. 1999,
  \apjl, 519, L97

\bibitem[\protect\citeauthoryear{{Brooks} \& {Warren}}{{Brooks} \&
  {Warren}}{2006}]{brooks2006}
{Brooks}, D.~H.,  \& {Warren}, H.~P. 2006, \apjs, 164, 202

\bibitem[\protect\citeauthoryear{{Cargill} \& {Klimchuk}}{{Cargill} \&
  {Klimchuk}}{1997}]{cargill1997}
{Cargill}, P.~J.,  \& {Klimchuk}, J.~A. 1997, \apj, 478, 799

\bibitem[\protect\citeauthoryear{{Cargill} \& {Klimchuk}}{{Cargill} \&
  {Klimchuk}}{2004}]{cargill2004}
{Cargill}, P.~J.,  \& {Klimchuk}, J.~A. 2004, \apj, 605, 911

\bibitem[\protect\citeauthoryear{{Culhane} et~al.}{{Culhane}
  et~al.}{2007}]{culhane2007}
{Culhane}, J.~L., et~al. 2007, \solphys, 60

\bibitem[\protect\citeauthoryear{{de Pontieu} et~al.}{{de Pontieu}
  et~al.}{1999}]{depontieu1999}
{de Pontieu}, B., {Berger}, T.~E., {Schrijver}, C.~J.,  \& {Title}, A.~M. 1999,
  \solphys, 190, 419

\bibitem[\protect\citeauthoryear{Feldman et~al.}{Feldman
  et~al.}{1992}]{feldman1992}
Feldman, U., Mandelbaum, P., Seely, J.~F., Doschek, G.~A.,  \& Gursky, H. 1992,
  \apjs, 81, 387

\bibitem[\protect\citeauthoryear{{Fletcher} \& {de Pontieu}}{{Fletcher} \& {de
  Pontieu}}{1999}]{fletcher1999}
{Fletcher}, L.,  \& {de Pontieu}, B. 1999, \apjl, 520, L135

\bibitem[\protect\citeauthoryear{{Korendyke} et~al.}{{Korendyke}
  et~al.}{2006}]{korendyke2006}
{Korendyke}, C.~M., et~al. 2006, \ao, 45, 8674

\bibitem[\protect\citeauthoryear{{Lundquist} et~al.}{{Lundquist}
  et~al.}{2004}]{lundquist2004}
{Lundquist}, L.~L., {Fisher}, G.~H., {McTiernan}, J.~M.,  \& {R{\'e}gnier}, S.
  2004, in ESA SP-575: SOHO 15 Coronal Heating, ed. R.~W. {Walsh},
  J.~{Ireland}, D.~{Danesy}, \& B.~{Fleck}, 306

\bibitem[\protect\citeauthoryear{{Martens}, {Kankelborg}, \&
  {Berger}}{{Martens} et~al.}{2000}]{martens2000}
{Martens}, P.~C.~H., {Kankelborg}, C.~C.,  \& {Berger}, T.~E. 2000, \apj, 537,
  471

\bibitem[\protect\citeauthoryear{{Mazzotta} et~al.}{{Mazzotta}
  et~al.}{1998}]{mazzotta1998}
{Mazzotta}, P., {Mazzitelli}, G., {Colafrancesco}, S.,  \& {Vittorio}, N. 1998,
  \aaps, 133, 403

\bibitem[\protect\citeauthoryear{{Peres}, {Reale}, \& {Golub}}{{Peres}
  et~al.}{1994}]{peres1994}
{Peres}, G., {Reale}, F.,  \& {Golub}, L. 1994, \apj, 422, 412

\bibitem[\protect\citeauthoryear{{Porter} \& {Klimchuk}}{{Porter} \&
  {Klimchuk}}{1995}]{porter1995}
{Porter}, L.~J.,  \& {Klimchuk}, J.~A. 1995, \apj, 454, 499

\bibitem[\protect\citeauthoryear{{Rosner}, {Tucker}, \& {Vaiana}}{{Rosner}
  et~al.}{1978}]{rosner1978}
{Rosner}, R., {Tucker}, W.~H.,  \& {Vaiana}, G.~S. 1978, \apj, 220, 643

\bibitem[\protect\citeauthoryear{{Schrijver} et~al.}{{Schrijver}
  et~al.}{2004}]{schrijver2004}
{Schrijver}, C.~J., {Sandman}, A.~W., {Aschwanden}, M.~J.,  \& {DeRosa}, M.~L.
  2004, \apj, 615, 512

\bibitem[\protect\citeauthoryear{{Schrijver} \& {van Ballegooijen}}{{Schrijver}
  \& {van Ballegooijen}}{2005}]{schrijver2005}
{Schrijver}, C.~J.,  \& {van Ballegooijen}, A.~A. 2005, \apj, 630, 552

\bibitem[\protect\citeauthoryear{{Vourlidas} et~al.}{{Vourlidas}
  et~al.}{2001}]{vourlidas2001}
{Vourlidas}, A., {Klimchuk}, J.~A., {Korendyke}, C.~M., {Tarbell}, T.~D.,  \&
  {Handy}, B.~N. 2001, \apj, 563, 374

\bibitem[\protect\citeauthoryear{{Warren} et~al.}{{Warren}
  et~al.}{2007}]{warren2007b}
{Warren}, H.~P., {Ugarte-Urra}, I., {Brooks}, D.~H., {Cirtian}, J.~W.,
  {Williams}, D.~R.,  \& {Hara}, H. 2007, \pasj, submitted

\bibitem[\protect\citeauthoryear{{Warren} \& {Winebarger}}{{Warren} \&
  {Winebarger}}{2006}]{warren2006b}
{Warren}, H.~P.,  \& {Winebarger}, A.~R. 2006, \apj, 645, 711

\bibitem[\protect\citeauthoryear{{Warren}, {Winebarger}, \& {Mariska}}{{Warren}
  et~al.}{2003}]{warren2003}
{Warren}, H.~P., {Winebarger}, A.~R.,  \& {Mariska}, J.~T. 2003, \apj, 593,
  1174

\bibitem[\protect\citeauthoryear{{Winebarger} et~al.}{{Winebarger}
  et~al.}{2002}]{winebarger2002}
{Winebarger}, A.~R., {Warren}, H., {van Ballegooijen}, A., {DeLuca}, E.~E.,  \&
  {Golub}, L. 2002, \apjl, 567, L89

\bibitem[\protect\citeauthoryear{{Winebarger} \& {Warren}}{{Winebarger} \&
  {Warren}}{2007}]{winebarger2007}
{Winebarger}, A.~R.,  \& {Warren}, H.~P. 2007, \apj

\bibitem[\protect\citeauthoryear{{Young} et~al.}{{Young}
  et~al.}{2003}]{young2003}
{Young}, P.~R., {Del Zanna}, G., {Landi}, E., {Dere}, K.~P., {Mason}, H.~E.,
  \& {Landini}, M. 2003, \apjs, 144, 135

\end{thebibliography}
\end{document}